The Electric Fields and "Lightning Jets" of the Sun and Solar Wind                10-Oct-2023

C. F. Driscoll,
Department of Physics,
University of California at San Diego
La Jolla, CA 92093 USA
Corresponding Author : cDriscoll@ucsd.edu

**Abstract**

A model of electric energization of the Solar Wind and Corona is developed, including electro-magnetic (EM) particle effects precluded by traditional magneto-hydro (MHD) assumptions.  Using standard 1-D radial Solar models for particle density and temperature, the Core gravito-electric field is calculated; and the range of possible Photospheric photo-electric fields is estimated.  The extant DC field apparently arises from about 460.Coulombs of charge displacement, mainly caused by the immense Solar energy flux pushing electrons outward.  Energetically, this electric field can accelerate surface protons out of the 2.keV gravity well and up to the 4. keV energies observed in the Fast Solar Wind.  The electrical energy is released in pervasive, persistent "proton lightning jets", which are proton beams, charge-neutralized by co-propagating electrons.  The jets are formed by pinched "avalanche breakdown" of the weakly ionized Photosphere, probably initiated on the down-welling edges of Solar surface granulations.  These energetic jets will glow as discrete filamentary surface spicules, and will be observed in reflect solar light as the diffuse K-Corona.   Significantly, the total charge displacement and observed Fast Wind Speed (FWS) are quantitatively predicted by a novel "virial limit": the (positive)  electrical potential energy at r=0 is limited in magnitude to the 10.keV gravitational well at r=0.

## I.  Introduction

This paper suggests a possible answer to one of the ESA/NASA mission "big questions", "What heats the Corona and energizes the Solar Wind?"  The 1-D radial model presented here shows that gravito-electric and photo-electric fields from the Solar Core plasma can be energetically sufficient to create the Fast Solar Wind of the quiet Sun.  The electric energy is released in pervasive, episodic "proton lightning jets" loosely analagous to Earth lightning.  The proton jet model is broadly consistent with observations of glowing "spicules" emanating from the solar surface, with Corona observations, and with satellite-measured characteristics of the Solar Wind.  That is, the electrically energized proton jets *are* the K-Corona and the Solar Wind.  Magnetic fields created by these (broadly neutral) charge currents can create the patchy, fluctuating magnetic fields observed on the Solar surface and in the Solar Wind.

Here, these electric charge effects are *calculated* for the Solar Core, and *estimated*  for the plasma sheath at the base of the Photosphere, based on standard 1-D radial models by Bahcall [4], Fontenla [21], and Avrett 2015 [2].  These models describe a near-static, near-equilibrium *fluid* state, with the strong Solar energy flux $\Gamma_\varepsilon \sim 64.\text{MW/m}^2$  of heat and light being the dominant flow; and no significant electric effects are explicitly included in the standard models.

A gravitating *plasma* of electrons and protons necessarily develops a "gravito-electric" ("Pannekoek") field $E_G$ in order to maintain 2-fluid force balance [41, 50, 18, 31, 55 ]. This DC field arises from $Q_G$ = 77.Coulombs of electrons displaced out of the Sun; here this charge and electric field is calculated directly from the standard-model [4] Core mass distribution. This minimal electric field $E_G$ is commonly incorporated into modern "exospheric" models of the Solar Wind [ 17, 32, 34, 43, 51-53].

Also, a confined plasma with strong outward electro-magnetic (EM) energy flow (photons) necessarily develops additional "photo-electric" polarization (here denoted $E_\gamma$ ) due to the outward "drag" of the energy flow on the (low-mass) electrons. This will be manifested as an additional charge $Q_\gamma$ displaced out of the plasma sheath, which is the (un-modeled) re-combination zone immediately below the Photosphere. Absent detailed knowledge of the EM/electron coupling, $Q_\gamma$ can not be quantitatively calculated, and it remains the only "unknown" in the simple electric model. We parametrize this by $\eta$ , as $Q_{tot} = Q_G + Q_\gamma = \eta\ Q_G$.

The resulting 1-D electric and gravitational potentials are then displayed over a broad heliospheric range, demonstrating that DC electric fields can plausibly accelerate surface protons out of the 2.keV gravity well, and up to a beam kinetic energy $\mathcal{E}_{pb}$ =4.keV (880.km/s); and this is closely equal to the maximum speeds observed in the Fast Solar Wind. [16, 36]

Significantly, this Fast Solar Wind energy is obtained from "maximal" charge ($\eta$ =6) which produces an electric potential at r=0 exactly as large as the Solar gravitational potential at r=0. This surprising broad "virial limit" provides strong support for the 1-D electrical model of Solar Wind energetics.

Even with established electric field energetics, the plasma dynamics of simultaneous proton and electron acceleration outward requires a nuanced understanding of "Ohm's Balance", described below. Most simply stated, the strong Solar energy flux $\Gamma\varepsilon$ "pushes" the electrons outward, and the electrons "drag" the protons with them, so as to maintain the total DC electric field.

In the electric model, the proton flow out of the Sun occurs in spatially pinched filamentary jets, due to localized "avalanche breakdown" of atmospheric resistivity, as in Earth lightning. Here, the breakdown is across the nominally neutral 2.Mm hydrogen atmosphere of the Photosphere, with a potential drop of ~15.Volts. These appear as glowing spicules, covering the entire Solar surface. In an "ideal" jet, protons from the Core plasma sheath are accelerated through the atmosphere along filamentary conduction paths, and then accelerated (with little drag) into the heliosphere.

Variations in spatial potential, neutral hydrogen entrainment, and interactions with other jet fields will produce a variety of jet profiles, including "partial launch", where the levitated protons remain gravitationally bound and return to the Sun. The patchy, intermittent "slow wind" observed near the ecliptic plane is probably caused by the jets interacting with ecliptic gas and dust.

The Solar jets are modeled as spatially pervasive and nearly continuous in time, with breakdown initiation points determined by the density and temperature variations of the convective cells (granulation) at the base of the Photosphere. For concreteness, the simple model posits $\sim 10^7$ extant proton jets, each pinched by $\sim 10^4$x, with lifetimes set by the convective cell lifetimes of $\sim 5$.minutes.  Initiation along the colder down-welling edges of cells seems likely [9], and is suggested by recent detailed Solar Orbiter images at UV wavelengths (10.eV), which often show exceptionally bright faculae [30], called "campfires" correlated with (and above) the cell edges (cool plasma). [40]  On larger scales of super-granulations [46] to sunspots, the surface structures may affect the electric and magnetic interactions among jets, leading to the stunning range of solar ejections,[8] from "levitated" prominences to coronal mass emissions.

The electrical model characterizes the electric energy and velocity of the jets, but not the magnitude of the proton flux; here, the plotted magnitudes are scaled to match satellite observations.  These proton jets propagating outward through the low Solar atmosphere will glow as "spicules", visible over 2-20 Mm. [45]  At moderate altitudes of $10^2$ - $10^3$ Mm, the keV-energy jets will reflect solar light as a diffuse K-Corona, replacing the venerable 1950's model of an "equilibrium, 200.eV electron gas"[26] with accompanying  protons. This "Core plasma sheath" origination of the Solar Wind is significantly different from the predominant "exospheric" models. [6]

These ideas of the electric model are developed in the next sections.  The discussion section describes connections and differences from some of the prior theories on the Corona and Solar Wind.

## II.  Standard Models

The standard fluid models of the Sun developed over the past 80 years provide a solid physics basis for the charged particle (plasma) effects proposed here.  The solar Core is well described by near-Local Thermal Equilibrium (LTE) equations; and the density and temperature profiles are tightly constrained by fusion physics and helioseismic observations.  [23]  In the Photosphere, Corona, and Heliosphere, the large energy flows "strain" the simple LTE concept of temperature, and empirical models are based on a plethora of  modern spectroscopic and in-situ particle and electro-magnetic measurements. Here, we describe the standard models and LTE fluid equations in broad overview, in order to identify the possible electric charge effects which are endemic to plasma "sheaths", as observed in Earth's atmosphere and in the laboratory.

Figure (1) displays the Bahcall model BS2005-OP [4] for the Core; and the Fontenla 1993 model [21] for the Photosphere.  Both models give radial profiles of baryonic particle densities $n_{tot}$, $n_{p+H}$ (black curves),  electron densities $n_e$ (black dashed), and temperature T (red).  Both models also estimate the energy diffusion length: as $\ell_T$ representing strong collisional coupling to T(r) in the Core; and as $\ell_\gamma$ representing bulk optical scattering, determined by matching extensive optical and spectral observations in the Photosphere. The derived curves labelled $E_T$, $E_\gamma$, and $\sigma_{\gamma e}$ will be discussed below.

The Bahcall Core model describes the Baryon ion density decreasing with radius from $10^{32}/m^3$ at r=0 to about $10^{27}/m^3$ at r=$R_{sun}$ −12.Mm =684.Mm. Here, $R_{sun} \equiv 696$.Mm is the bottom of the Photosphere, defined by Optical Depth equal to 1. The model includes detailed estimates of hydrogen, helium as well as carbon, nitrogen and oxygen distributions, for accurate representation of fusion energy release and neutrino creation. The resulting temperature varies from a central 1.35 keV ($1.57 \times 10^7$ Kelvin) to 7.0 eV at r=684.Mm. Over this range, the species ionization is fully predictable and almost complete.

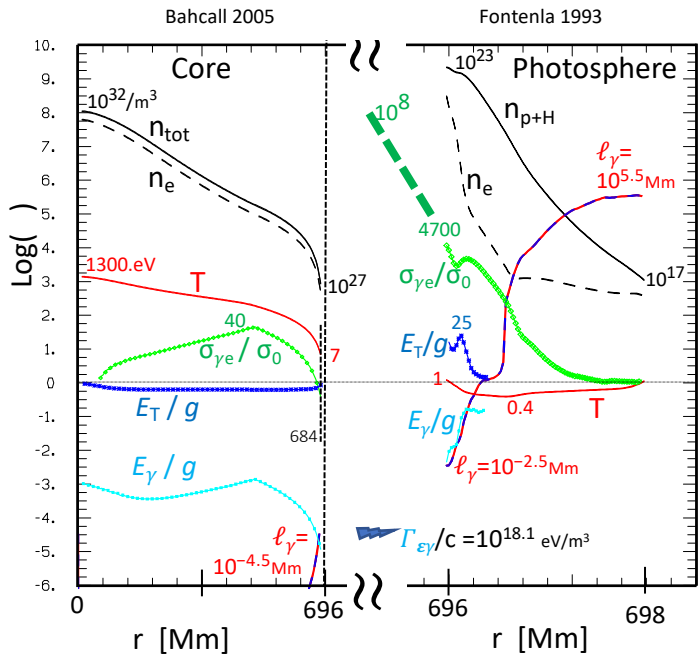

Figure 1/ Standard Solar models for the Core and Photosphere (with expanded radial scale), presenting particle densities n, temperature T, and optical absorption length $\ell_\gamma$. From these, the cross-section $\sigma_{\gamma e}$ and electric field contributions $E_T$ and $E_\gamma$ are estimated from Eqns(5-7).

The Fontenla Photosphere empirical model is for a hydrogen-only star scaled to the Sun, thereby simplifying the ionization and recombination effects. The model is based on spectroscopic measurements of photon absorption and emission, and so tabulates an optical absorption length $\ell_\gamma(r)$. The model begins 0.1 Mm below OD=1 at r=696.Mm, and extends somewhat past r=698.Mm. (Note that the term Chromosphere is often used for the region above OD=1). Over the central range displayed, the best-fit temperature is nearly constant at 0.4 eV; and the atmosphere is a barely ionized hydrogen gas, with ionization ranging from 0.03 down to $10^{-4}$ then back up to 0.3. Beyond $R_s$+2.Mm, the modeled temperature rises abruptly to form a hot Corona, as discussed with Fig(5).

## A. LTE Fluid Equations

Here, we sketch the basic fluid equilibrium equations used in solar models [11], extended to include the electric forces on the separate proton and electron fluids. To simplify discussion of physics effects, only the Hydrogen species is considered; "quasi-neutrality" $n_p \sim n_e$ is assumed; and both gravity on electrons and EM wave coupling to protons are neglected. Also for simplicity of comparison, energies in eV are displayed rather than potential energies, as $\Phi \equiv e\phi$ , $\Psi_G \equiv m_p \psi_G$ ; and $T \equiv k_B T_{Kelv}$ ; and forces are in eV/m, as $E \equiv -\Phi'$, $g \equiv -\Psi_G'$. In simplest form, the 2-fluid equations are:

$$\nabla^2 \Psi_G(r) = (4\pi G\, m_p^2)\, n_p(r) \qquad (1a)$$

$$\nabla^2 \Phi(r) \;\; = -(e^2/\varepsilon_0)\,(n_p - n_e) \qquad (1b)$$

$$\nabla \cdot \Gamma_\varepsilon(r) = \frac{d}{dt}\, \mathcal{E}_{vf}(r) \qquad (2)$$

$$(\sigma_{SB}\, T^4)'\, l_T = -\Gamma_\varepsilon \qquad (3)$$

$$[n_p T]' \qquad\qquad + n_p \Psi' + \;\; n_p \Phi' = 0 \qquad (4a)$$

$$[n_e T]' \;\; - \frac{\Gamma_{\varepsilon\gamma}}{c\, l_{\gamma e}} \qquad\qquad - n_e \Phi' = 0 \qquad (4b)$$

$$\big[(2n)\, T\big]' - \frac{\Gamma_{\varepsilon\gamma}}{c\, l_{\gamma e}} + n\,\Psi' \qquad\quad = 0 \qquad (4a + 4b)$$

$$[\delta n_{pe}\, T]' + \frac{\Gamma_{\varepsilon\gamma}}{c\, l_{\gamma e}\, n_e} + \Psi' + \;\; 2\Phi' = 0 \qquad (4a - 4b)$$

$$\big\{ n_{pb}\, T \big\}' \qquad\qquad + n_{pb} \Psi' + n_{pb} \Phi' \;\triangleq\; (n_{pb}\, m_p\, v_{pb}^2)' \qquad (4a^*)$$

Eqns(1a,1b) are "parallel" Poisson equations, but the electrostatic interaction is about $10^{36}$ times stronger, and $\delta n_{pe} = (n_p - n_e)$ will be approximately $10^{-36}\, n_p$ . The volumetric fusion energy production of Eqn(2) drives the large energy flux $\Gamma_\varepsilon$ , and in the high density, collisional Core this is modeled as Stefan-Boltzman radiative equilibrium with temperature variation length $l_T$ of Eqn(3). Eqns(4a,4b) represent separate force balances for protons and electrons, with no net charge current and hence no p-e coupling forces.

The $\Gamma_{\varepsilon\gamma}/c\, l_{\gamma e}$ radial force density on electrons (often ignored) arises from the net-outward momentum of EM radiation which is scattered by electrons. For each electron, this can be expressed as a cross-section $\sigma_{\gamma e} = 1/n_e\, l_{\gamma e}$ ; but this cross-section varies by factors of $10^8$ depending on the electron *correlation* with nearby protons, as will be discussed below.

The sum in Eqn(4a+4b) gets back to the more common 1-fluid perspective, where the electric energies "disappear". The difference in Eqn(4a-4b) exposes the electric energy, especially when $\delta n_{pe}$ and $\Gamma_{e\gamma}$ are negligible. Terms in Eqn(4a*) will be discussed below in relation to proton beam energies in Fig(3).

Integrating any version of the stellar equilibrium equations is a difficult and "artful" process, especially for the Photosphere; and no such integration is attempted here. Rather, Eqns(4) are applied to the standard non-electric models, to *estimate* non-LTE contributions to the electric field, when one term or another dominates.

### III. Non-LTE Electric Effects

The description of Local Thermal Equilibrium necessarily fails when there are large energy flows or strong spatial gradients, such as at the plasma edge; and in these edge "sheaths" the electric fields generally become dominant [25,54]. Here, we define and discuss the 3 separate electric force magnitudes driven by the 3 separate terms in Eqns(4a) and(4b), representing thermo-electric, photo-electric, and gravito-electric effects. Each effect is strictly valid only when the other two are negligible, and the meta-equilibrium of the Sun is apparently a tightly inter-linked balance of competing effects.

**( Gravity)** The (Pannekoek) gravito-electric effect is simple, and broadly applicable to the Solar models. [41, 50, 18, 31, 55] For the high-density Core where the "photon drag" (per electron) is negligible, Eqn(4a-4b) gives

$$E_G(r) \equiv -(1/2)\,g(r)$$
$$= -(1/2)\,(2.8\,\text{eV} / \text{Mm})\ @\,R_s \tag{5}$$

Here, the $E_G = -\boldsymbol{\Phi}'$ (outward) electric force counteracts half of the $g = -\boldsymbol{\psi}'$ gravity force on every proton. Of course, $E_G$ produces an opposite (inward) force on electrons, keeping them contained in the Sun. This electric field is often ignored, since it sums to zero on an "un-charged" fluid, but the associated Solar charge $Q_G$ will be discussed in the next section.

**( Thermal)** Without gravity and EM energy flows, Eqns(4a) or (4b) would represent the thermo-electric effect, as

$$E_T(r) \equiv \frac{[nT]'}{n}. \tag{6}$$

This expresses the ubiquitous entropic/electric pressure balance, which reduces to $ne\varphi = nT$ for a confined single charge species.

For the Core, direct calculation of Eqn(6) from the density and temperature profiles gives $E_T(r)/g(r) \sim 0.6$, displayed as $\text{Log}_{10}(E_T/g) \sim -0.22$ in Fig(1). Summed together with $E_g(r)/g(r) = 0.5\,g(r)$ (not plotted) the sum approximates the proton equilibrium force-balance of Eqn(4a).

For the Photosphere, calculation from the model profile shows a thermal field $E_T$ peaking at $E_T/g \sim 25$ near $R_s$, suggesting that protons would be levitated by 25x gravity. This will be discussed in relation to $E_\gamma$ next.

**( Photo $\gamma$)** The non-LTE "photo-electric drag" force on electrons can be expressed as

$$E_\gamma(r) \equiv \frac{\Gamma_{\varepsilon\gamma}}{2c\, l_{\gamma e}\, n_e} \equiv \frac{\Gamma_{\varepsilon\gamma}}{2c} \sigma_{\gamma e}$$

$$= (46.\mu\text{eV/Mm}) \, (\sigma_{\gamma e}/\sigma_0) \, (r/R_s)^{-2} \qquad (7)$$

$$\leq (4.6\,\text{keV/Mm}) \text{ for } \sigma_{\gamma e}(\text{H}^-)$$

Here, the EM radiation energy flux $\Gamma_{\varepsilon\gamma}$ is taken to be dominant compared to conductive and convective flows. However, the photon-electron cross-section $\sigma_{\gamma e}$ is highly variable and difficult to evaluate: theory describes the minimal "Thompson" cross-section for an *isolated* electron as $\sigma_0 = 0.67 \times 10^{-28}\, m^2$ ; but $\sigma_{\gamma e} \sim (10^4 - 10^8)\,\sigma_0$ in regions of electron-proton *correlation* , i.e. plasma recombination. [58]

For the standard Core model, the slow outward energy transport corresponds to a small collisional $\ell_T$ in Eqn(3). If this were interpreted as mainly photon energy transport with $\ell_{\gamma e} = \ell_T$, then Eqn(7) would give a negligibly small $E_{\gamma \ll E_T}$, and would give a $\sigma_{\gamma e}/\sigma_0$ varying from 1 to 40, as shown in Fig(1). The peak at 40 is a weak "trace" of thermal conductivity becoming convective at 0.7 $R_s$ .

In contrast, the empirical Fontenla Photosphere model tabulates both $\ell_{\gamma e}$ and fractional ionization $n_e(r)$ at "equilibrium" temperature T(r). The calculated $E_\gamma$ is seen to peak at approximately 0.5, while $\sigma_{\gamma e}/\sigma_0$ is as large as 4700 near $R_s$. However, both estimates are overly sensitive to the $n_e \ell_\gamma$ product, and are oscillatory above 0.5Mm (not plotted).

Unfortunately, Figure (1) leaves an un-modelled "recombination zone" of 12.Mm immediately below $R_s$, where the plasma state evolves from fully ionized to almost fully re-combined; where binding energies (e.g. 13.6 eV) are released; and where the particle couplings $\sigma_{\gamma e}$ to the heat flux $\Gamma_\varepsilon$ are large and difficult to estimate.[28,37] In this zone, the modeled bulk optical absorption length varies over $10^{-4.5} < \ell_\gamma < 10^{+4.5}$ Mm, so neither collisional nor collision-less equations apply well.

It is here that the new electrical model posits strong photo-electric coupling and charge displacement. The photo-electric coupling strength is represented schematically by the heavy dashed green line extending the normalized $\sigma_{\gamma e}$ from the modelled 4700 up to the $10^8$ range expected in the recombination zone. This strong "photon drag" maintains a steady electron displacement outward, generating the electric field $E_\gamma$, which adds to the Pannekoek electron displacement and field $E_G$ in the Core.

Moreover, the "virial limit" energetics of the electric model suggest that $E_G + E_\gamma \sim 3g$ for $r \gtrsim R_s$, generated by a steady electron displacement of $Q_{tot} \sim 6\,Q_G$. This will be discussed next.

## A. Charge Displacement, 1-D Energetics

The electric model quantitatively calculates the gravitational force g(r) from the tabulated Core mass distribution and the Poisson Eqn(1a), constrained by the total mass $M_s = 1.989\mathrm{x}10^{30}$ kg $= 10^{57.08}$ m$_p$. The corresponding "energy well" depth $-\psi_G(r)$ is plotted in green in Fig(2), being 10.keV at r=0 and 2.keV at r=R$_s$, and falling off as $r^{-1}$ for r>R$_s$.

The required Pannekoek electric force $E_G(r)$ is then calculated from Eqn(5), and the corresponding displaced charge Q$_g$(r) is plotted in blue in Fig(2), integrating to

$$Q_G = 77.\mathrm{Coulomb} = 10^{20.68}\,\mathrm{e}\,. \qquad (8)$$

The corresponding electric energy $\Phi^{(1)}(r)$ is plotted in red, with superscript (1) signifying that $\eta$=1 and Q$_G$ is the only charge displaced from the Sun. The Pannekoek-only electric energy is then 1/2 of $-\Psi_G(r)$, being 5.keV at r=0 and 1.keV at r=R$_s$, and also falling off as $r^{-1}$ for r>R$_s$. That is, the 1.keV Pannekoek-only electric energy is insufficient to eject surface protons from the 2.keV Solar gravity well, and modeling an (un-charged) "exosphere" can not improve the electric/gravity force ratio.

Here, the electric model posits that strong photo-electric effects displace an additional electron charge Q$_\gamma$ out of the "recombination zone" around $r \sim R_s$. This additional "plasma sheath" charge displacement is quantified by the one adjustable model parameter $\eta$, as $Q_\gamma = (\eta - 1)\,Q_G$, or

$$Q_{tot} = \eta\,Q_G\,. \qquad (9)$$

Figure (2) displays the calculated Solar electric energy $\Phi^{(\eta)}(r)$ for four illustrative possibilities, $\eta = \{1,2,4,6\}$, with the the additional charge Q$_\gamma$ represented by blue arrows at R$_s$.

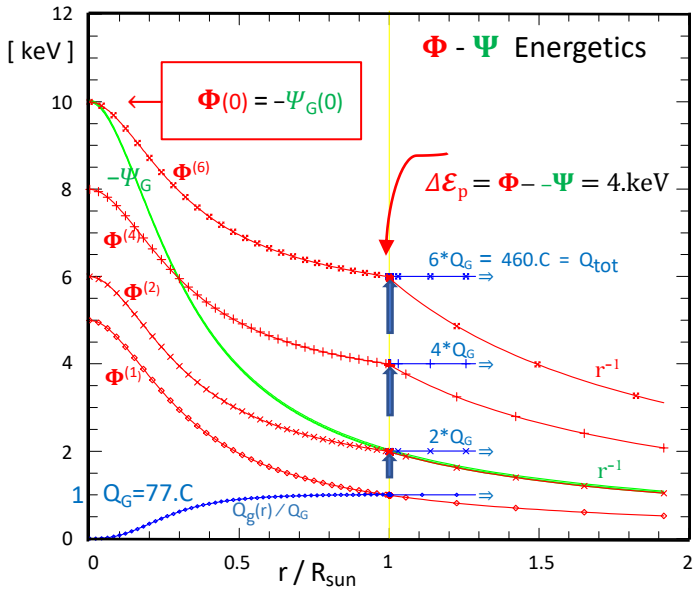

Figure 2/ Radial profiles of electric potential determined from the calculated gravito-electric charge $Q_G$ distributed over $0 < r < R_s$, plus 4 choices for additional photo-electric charge $Q_\gamma$ localized to $r = R_s$ (blue arrows), giving $Q_{tot} = \eta \cdot Q_G = \{1, 2, 4, 6\} \cdot Q_G$ . Energy is available for proton acceleration when $\Phi > -\Psi_G$ for $r > R_s$ , i.e. when $\eta > 2$ .

## B. Maximal 1-D Proton Kinetics

Figure (2) shows that for $\eta > 2$ , the (positive) electric potential at $R_s$ is greater than the (negative) gravitational potential, giving a positive energy $\Delta\mathcal{E}_{pb} = \Phi(R_s) - (-\Psi_G(R_s))$ to accelerate surface proton beams out of the gravitational well. In this simple model of (Pannekoek) Core-distributed charge $Q_G$ , plus $R_s$-localized charge $Q_\gamma = (\eta-1)Q_G$, the available energy $\Delta\mathcal{E}_{pb}$ depends only on $\eta$, and on the radial distribution of the Solar Core mass. The Core mass distribution can be characterized by $Z \equiv \Psi_G(0) / \Psi_G(R_s)$ , with a calculated $Z = 5.0$ for the Bahcall model. Defining
$$\mathcal{E}_G \equiv -\Psi_G(R_s) = 2.\text{keV} , \qquad (10)$$
this simple mass and charge distribution predicts :
$$\Phi(0) = (Z + \eta - 1)\,\mathcal{E}_G, \text{ and} \qquad (11)$$
$$\Delta\mathcal{E}_{pb} = (\eta-2)\,\mathcal{E}_G\,/2 . \qquad (12)$$

Significantly, for $\eta = 6$ , the electric potential $\Phi$ at r=0 is as large as the gravitational potential $-\Psi_G$ at r=0. This "maximal" electric potential arises from a total charge displacement $Q_{tot}^{max} = 463.\text{C} = 10^{21.5}$ electrons, out of $10^{57.1}$ total electrons in the Sun. The surface electric force is $E^{(max)} = 8.6\text{eV/Mm}$ , closely equaling 3 times gravity on a proton at $R_s$ . When a surface proton is accelerated outward with no intervening dissipation, the asymptotic ($r \gg R_s$) kinetic energy is $\Delta\mathcal{E}_{pb}^{max} = 4.\text{keV}$, giving an asymptotic proton velocity of $v_{pb}^{max} = \text{sqrt}(2\mathcal{E}_{pb}^{max}/m_p) = 880.\,\text{km/s} . \qquad (13)$

This maximal proton speed is about 10% greater than the persistent and pervasive Fast Wind speed (~800.km/sec) which is observed by satellites (particularly Ulysses [36]) travelling out of the ecliptic plane.

This strongly suggests that the central electrical potential is limited to the central gravitational potential, in a manner similar to the Pannekoek relation; and that this determines Fast Wind proton energy.  The observed consistency over years, over (non-ecliptic) latitudes, and over radius also strongly suggests a global connection between gravity and electric energetics.  The local Pannekoek relation and approximate thermal connections [28,37] are sometimes discussed in relation to the "virial theorem" of dynamics; but this surprising global "virial limit" has apparently not been observed or vetted in other situations.  In terms of dynamics, it is probable that the $\eta<6$ state would be unstable towards increased $Q_\gamma$, so $\eta=6$ may be both a lower limit and an upper bound, i.e. the uniquely specified state.

Figure (3) displays a global perspective of this maximal 1-D electric potential driving a proton wind, with dissipationless kinetics.  Here, the difference between $\Phi(h)$ and $-\Psi_G(h)$ is the kinetic energy $\mathcal{E}_{pb}(h)$ , asymptoting to $\Delta\mathcal{E}_{pb}=4.keV$.

The $\mathcal{E}_{pw}$ and $v_{pw}$ curves are radial integrals of Eqn (4a*) with $\{n_{pb}T\}'=0$.  Here, the only energy is the difference between the electrostatic potential $\Phi$ and the gravity well depth $-\Psi$ , and this all goes to the kinetic energy of the proton beam.  The plotted solution starts from a small energy $\mathcal{E}_0=10.eV$ at h=0, thereby bypassing the dynamics of wind formation from "runaways" or "avalanch breakdown", as discussed in the next two sections.  The most rapid proton energy gain is between $h=10^2$ and $h=10^3$ Mm,  due to the $R_s=696$ Mm scaling of the spherical global potentials.

For comparison, the 4 blue dots labelled $O^{VI}$ are the radial H0 velocities  u  from the self-consistent empirical model A2 of Cranmer 1999 [14], for the quiet Sun over a polar Coronal hole.  The model is based on SOHO UV Coronagrah Spectrometer measurements of H0 and $O^{5+}$ lines.  The similar Model A1 has about 30% lower velocities.  Also shown are the persistent out-of-ecliptic Ulysses velocity (blue U); and the central range of in-eclliptic velocities measured at 1AU by the ACE satellite (blue A).

In Figure(3) (and Fig(5) below) , the average proton beam density  $n_{pb}(r)$ and flux  $n_{pb}$ $v_{pb}(r)$  are scaled in magnitude to match the extensive satellite measurements of the Fast Solar Wind flux,  which scale accurately as  $r^{-2}$ , representing near complete particle conservation in the constant-velocity flow.   That is, the electrical model does not determine the magnitude of the flow density, but flow conservation is implicit in Fig(3).

In Figure(3) (and Fig(5) below) , the average proton beam density  $n_{pb}(r)$ and flux  $n_{pb}$ $v_{pb}(r)$  are scaled in magnitude to match the extensive satellite measurements of the Fast Solar Wind flux,  which scale accurately as  $r^{-2}$ , representing near complete particle conservation in the constant-velocity flow.   That is, the electrical model does not determine the magnitude of the flow density, but flow conservation is implicit in Fig(3).

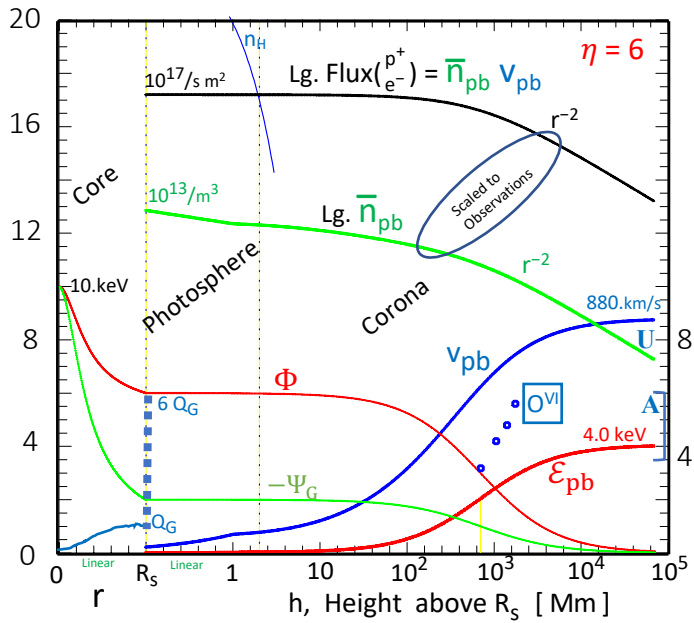

Figure 3/ Global perspective showing $\eta=6$ "maximal" charge Q profile in the Core (blue) asymptoting to $Q_G$, , and 5 times as much additional charge in the plasma sheath located at $r\leq R_s$ (blue squares), giving $\Phi=5.$keV and E=8.6eV/Mm just above the sheath . The energy difference $\mathcal{E}_{pb}$ between $\Phi(r)$ and $-\Psi_G$ accelerates protons to wind speed $v_{pb}$ . When scaled in density magnitude to match satellite observations, this gives the average proton wind density and flux profiles shown on a $Log_{10}$ scale.

Parker's "supersonic hydrodynamic wind" solution was obtained from Equation (4a*) with the $\{n_{pb}T\}'$ term included,  the $n_{pb}\Phi'$ term deleted,  and the assumption of "constant Temperature" and "heat conduction".  [42,10].  This suggests that Coronal thermal energy can somehow lift protons out of the 2.keV gravity well and up to keV energies.  Parker's solution was criticized by Chamberlain  [12] as mathematically  invalid, but it became the "default", because no other solution obtained Fast Wind energies near 4.keV; and it is now treated with reverence. [ 24]

Section IV  addresses the Ohm's Balance question of, "How does the DC electric field cause both positive protons and negative electrons to flow outward ?"  In simplest terms: "The hellacious EM energy flux of 64.MW/m² preferentially pushes light electrons outward, and the electrons drag the protons with them, as they must to maintain a static (ambipolar) electric field."

A second perplexing question is, "What is the effect of the resistive, almost-neutral Hydrogen atmsphere bounding the polarized plasma sheath at the bottom of the Photosphere?  The most direct answer is, "As with Earth lightning, charge displacement will build up until localized avalanche breakdown occurs, forming localized charge Jets."

In the Photosphere, the surface-averaged charge displacement is constant, and the lightning jets are persistent and pervaise across the Solar surface.

## IV. Ohm's Balance, Runaways, Proton Lightning Jets

"Ohm's Balance" in plasmas is substantially more complex than "Ohm's Law" in resistive wires and laboratory plasmas [22], as shown schematically in Figure(4). Here we describe different "parts" of the total electric field, identified by their causes.

a) In wires, an "external" electric force $E_{ext}$ on the electrons is balanced by the average collisional drag on the electron flow. This average drag is equivalent to a reverse electric force $<E_{coll}>$, with $E_{ext} = -<E_{coll}>$ in steady flow.

b) In the collisional Core of the solar plasma, the electrons see a balance between the entropic thermal gradient force $T'$, and the static inward gravito-electric force $E_G$ arising from the few displaced electrons leaving "un-neutralized" protons behind. This displaced charge integrates to 77.C, giving $E_G \sim 1.4$ eV/Mm = g /2 just below r=$R_s$ .

The protons see an outward gravito-electric force $E_G$ as counteracting ½ the force of gravity g , with the the thermal gradient force $T'$ countering the other half of gravity.

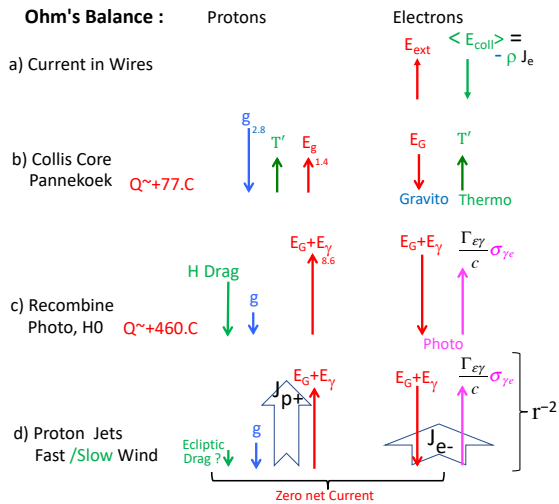

Figure 4/ Ohm's balance of forces on protons and electrons, in 4 regimes: (a) for electron current in wires, where colisions create a fluctuating drag force $<E_{coll}>$ ; (b) for the collisional Solar Core, where the Pannekoek field $E_g$ contains the electrons; (c) for the recombination zone, where the photon-electron cross-section $\sigma_{\gamma e}$ is dominant; and (d) and for an avalanche breakdown proton jet.

c) In the low-temperature "recombination zone" and lower Photosphere, the electrons have a large cross-section $\sigma_{\gamma e}$ for the outward solar energy flux $\Gamma_{\varepsilon \gamma}$ , and a larger force $E_G + E_\gamma$ is required for an equilibrium electron balance. Here, this implicitly includes the entropic thermal force.

The protons then experience a net upward electric force $E_G + E_\gamma$ which can be greater than gravity, here shown as the virial "limit" of $E_{tot} = 8.6$ eV/Mm = 3g. However, the 2.Mm Photosphere is basically dense neutral hydrogen, which tends to limit the proton flow through it. This *could* result in a slow upward proton drift; or the protons could develop a weak "runaway" fraction; or filamentary "proton jets" could develop through an "avalanche breakdown" of flow resistivity, similar to Earth lightning.

d) In the electric proton jets of the Solar Wind, protons are accelerated outward by $E_g + E_\gamma$ which is 3g from "maximal" charge $\eta$=6, less drag from background plasma, gas, and dust. This drag is apparently most significant in the ecliptic plane [36], as discussed below.

The low-mass electrons require a negligible force to accelerate to proton speed, since electrons at 1eV have a thermal speed ~400.km/s, comparable to Solar Wind speeds. However, this requires a close force cancellation between inward $-(E_g + E_\gamma)$ and the outward "push" from the energy flux $\Gamma_{\varepsilon\gamma}$. This close cancellation probably involves the dynamics of electron/proton correlation (i.e. near recombination) in the plasma jet: when an electron "cools" towards recombination, its cross-section $\sigma_{\gamma e}$ increases exponentially, and it is "heated" away from recombination by the photon flux.

The electrons provide broad charge neutralization and current cancellation, although transverse dispersion would result in local currents and charges. The major forces (except drag) scale as $r^{-2}$, as does the p+/e- particle flux density.

## A. 1-D Runaways

Under some circumstances, an upward electric force on gravitationally confined protons could result in some energetic particles (runaways) emerging out of a "viscous" neutral atmosphere. However, the avalanche breakdown (a.k.a. lightning) effect described next is substantially more effective.

The Dreicer runaway effect [44] occurs for "Maxwellian tail" particles (here protons) when the electric energy gained $E_D \ell_s$ in a collision length $\ell_s$ is greater than the average particle energy lost per collision $\mathcal{E}_p$ with background species "s". Since the collisional cross-section $\sigma_{ps}$ generally decreases with particle velocity, the particle may accelerate to ever-higher velocity. This can be approximated as $E_D \ell_s \gtrsim n_s \sigma_{ps} \ell_s \mathcal{E}_p$ .

Figure (5) displays a simple PIC simulation which illustrates the runaway effect, which quantifies the limited applicability to the Photosphere, and which illustrates similarities to "avalanche breakdown". At z=0, a Maxwellian velocity distribution (0.5eV) of protons is accelerated against gravity (2.8eV/Mm) by a constant 8.6eV/Mm electric field arising primarily from the photon energy flux $\Gamma_\varepsilon$. The neutral Hydrogen density $n_H$ decreases exponentially, to $10^{14}$/m$^3$ at z=1, and $10^{11}$ at z=2. The proton-H scattering cross-section[29] is approximated as $\sigma_{pH} \sim (28 \times 10^{-20}$ m$^2$ ) $\mathcal{E}_p^{-1}$ . A beam of accelerating protons emerges at z~1.5, and accelerates to $v_p$ = 65.km/s at z=4.

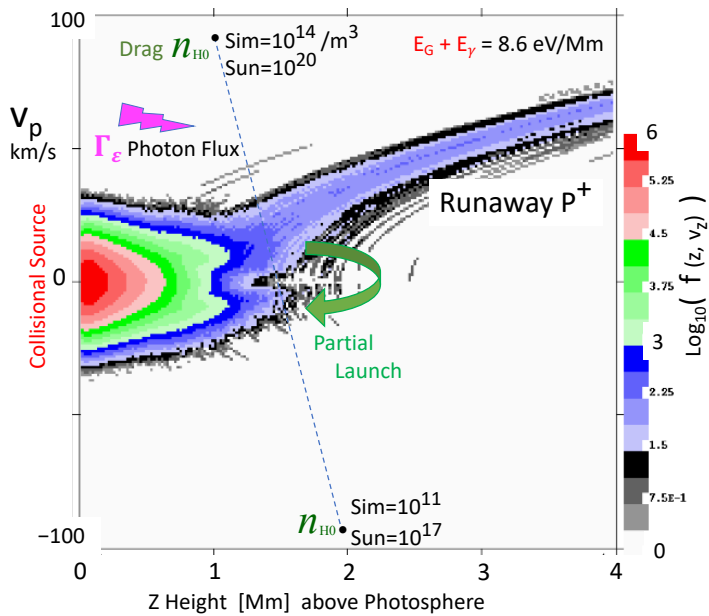

Figure 5/ 1-D schematic illustration of runaway protons emereging from a collisional thermal atmosphere, with decreasing neutral Hydrogen background gas drag.

This simulation incorporates background drag approximately 10$^6$ times *weaker* than expected for the Photosphere, by virtue of $n_H$ being specified as 10$^6$ times lower at any z-position. With the unrealistically low simulation density, the z=1 Dreicer electric field is $E_D \sim 28$. eV/Mm , and runaway occurs as shown. In contrast, the standard models show H density above 10$^{17}$ for z<2, making the runaway effect 10$^3$ to 10$^6$ times less effective. That is, the collisional Hydrogen atmosphere would appear quite resistive compared to the weak fields being posited here. Earth's atmosphere would appear even less prone to runaway electrons, but lightning occurs in a wide variety of regimes, including upward into the mesosphere.[47]

Figure (5) does illustrate another important effect, namely "partial launch", visible as the lower "stubbed" tail of returning particles, which have barely missed the transition to an escape trajectory. With random episodic collisional decelerations, some energetic protons become gravitationally re-bound at low velocity, and then return to z=0, depositing their kinetic energy along the way. Similar randomization of electric energy may occur in lightning jets, discussed next.

**B. Avalanche Breakdown,  Lightning Jets**

In an avalanche breakdown of resistivity, energy is released when the initial accelerating particles heat the background gas/plasma, lowering the resistivity along a specific path, thereby causing exponentially increasing flows and heating.  Earth lightning is a thoroughly observed exemplar of this process, demonstrating pinched (and branching) paths with apparent length/diameter ratios $\sim 10^2$ to $10^3$ .

Here, we propose that DC electric field energy is released as proton acceleration and plasma heating, along localized breakdown paths through the hydrogen gas Photosphere. This produces filamentary "Proton Lightning Jets" of energetic protons, accompanied by less well-focused electrons in overall equal numbers.  The discrete filamentary jets are created by "current pinch" effects which favor small diameter particle currents (as with Earth lightning) rather than spatially broad flows (as in a glow discharge).

Here, each proton Jet is (somewhat arbitrarily) modeled as $\sim$(5.km)$^2$ in area, one Jet per convection cell occupying (0.8Mm)$^2$ , giving $\sim 10^7$ Jets over the solar surface at any given time.  Due to excitation of background gas and plasma,  these will appear as pervasive, isolated spicules in the Photosphere, typically visible to a height of 4 - 8 Mm, with some visible up to 10-20 Mm in the UV and EUV. [45]

Figure (6) shows the radial density profile in these Lightning Jets and average Wind speed (beaded lines) , in relation to the "semi-empirical" Avrett 2015 model [2] of the extended Photosphere; and to the 4 traditional static "electron gas" atmosphere models of the K-Corona,  [3, 20, 56, 14]  here shown in order of increasing starting height of 35 - 500.Mm . The electric model is displayed with the "maximal" $\eta$=6.

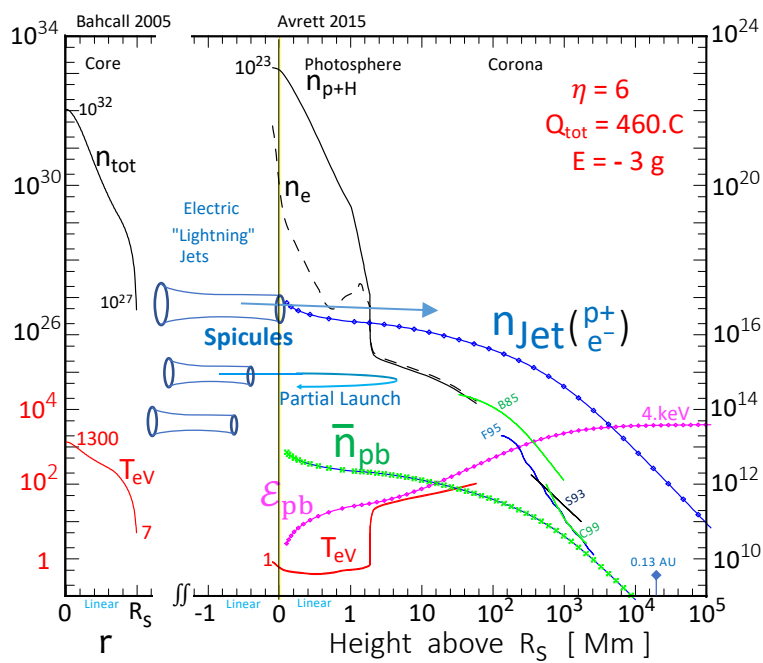

Figure 6/ 3-D pinched "Proton Lightning Jets" emerge from the lower Photosphere, glowing as Spicules. From (2–50)Mm, the protons gain energy $\mathcal{E}_{pb}$ ~(15–300)eV, somewhat above the semiempirical model $T_{ev}$, eventually asymptoting to 4.keV. The average proton/electron beam density $n_{pb}$ is quantitatively determined by $E_{pb}$ together with the measured fast wind proton flux, and 4 traditional K-Corona models merge with $n_{pb}$ at 2000.Mm

The empirical Avrett model [2]extends the Fontenla 1993 model[21] of Fig(1) to include extensive observational data and atomic properties for 32 atoms and ions; and also extends the fits to a height of 58.Mm. At a height of 58.Mm, the Avrett temperature fit has risen to 100.eV, and the density has dropped to $1.4 \times 10^{14}$. The electric model predicts proton beam kinetic energy $\mathcal{E}_{pb}$ rising at 5.6eV/Mm , in general agreement with the empirical temperature fit shown as $T_{eV}$. That is, the electric energy is more than sufficient to create a "heated" corona.

Figure(6) also displays proton/electron densities from 4 traditional K-Corona models, based on Solar light scattering by (un-correlated) coronal electrons with $\sigma_{\gamma e} = \sigma_0$. These are Badalyn 1985 [3], Fisher 1995 [20], Strachan 1993 [56], and Cranmer 1999 [14]. These model densities merge with the quantitatively determined $n_{pb}$ at a height of 2000.Mm ($\sim 4.R_s$)

In addition to the outward beam, there may be "partial launch" protons and electrons from the beam itself, as well as other in-flowing gas and plasma. At present, the electric model gives no insight into the magnitude of these partially levitated and returning particles. Some beams may have barely missed the transition to escape velocity, due to global variations in electric fields, or in the plasma recombination state at the source, or in the amount of neutral Hydrogen entrainment.

Of course, in the electric model, these effects depend on the charge, mass, and ionization potentials of the various elements besides hydrogen. The ionization of He and CNO "metals" is crucial in determining the electron fraction in the Photosphere; and exceeding low "first ionization potential" elements such as Lithium ( FIP =5.6eV ) may behave exceptionally in stellar electric fields.

The pinched Jet model is consistent with Solar images at UV wavelengths (10.eV), which show energetic "bright points" (a.k.a campfires) along the cool down-welling edges of the ubiquitous surface convection cells [9], suggesting that this avalanche breakdown occurs most readily there. These convection cells typically remain individually coherent for ~5.minutes, similar to the average lifetime of the observed bright points. It is likely that the spatiality and longevity of the massively energetic convection cells are dominant in determining the locations and lifetimes of the "puny" Lightning Jets.

The Solar Wind flux at $R_s$ is about $10^{17}$/s $m^2$ , so each convection cell jet would carry a proton flow of about ¼ $\times 10^{29}$/s , with broadly accompanying electrons. This represents $4 \times 10^9$ Amps of each species, giving magnetic field magnitude 0 < B < 0.3 Tesla at 2.5km from the Jet, depending on the dynamical "neutralization" or "return current" placement. Statistical and dynamical agglomeration into individual jets into larger structures could create magnetic fields of greater strength and extended multi-pole structure.

**V. Discussion**

The quantitative electric model presented here differs from prior Solar Wind models in 3 broad aspects. First, plasma *electro*-dynamics is analyzed, without the restrictions and assumptions of ideal MHD. [1, 13, 42] Electric energies are presumed to be significant and to drive currents, which create magnetic fields which may also affect the dynamics. Second, the approach is one of plasma particles, rather than continuous fluids and gases. Here, all kinetic, thermal, electro-magnetic and gravitational energies are expressed in particle-relevant electron-Volts, including the immense EM energy flux. Third, the Solar plasma is considered as a whole, including the plasma *sheath*; and real-world sheaths necessarily involve electric field, localized charge, and the ejection of particles.

Sheaths are complicated and non-linear, presenting unsolved puzzles even in the laboratory with in situ diagnostics from probes and laser-induced fluorescence. [25] With multiple ion species, [54] the sheath exhibits differing pre-acceleration for each species, modifying the simple Bohm velocity prediction.

In the electric model, the solar wind protons emerge from the recombination sheath between the Core and Photosphere, driven by the electric field from the sheath "polarization" charge displacement. This might also be called an "ambipolar" electric field. The overlying 2.Mm neutral hydrogen atmosphere is ostensibly resistive to proton flow, but a pinched avalanche breakdown path enables (and perhaps regulates) the proton jet flow.

The electric model describes these "Lightning Jets" as spatially localized beams of plasma, with momentum determined by the massive protons, and charge neutralization from low-mass electrons. Such plasma jets can travel in straight lines even in the presence of a moderate magnetic field B, by developing an internal polarization (i.e. Hall) field $E_{pol} = v_{Jet} \times B$ .

Thus, plasma jets propagating through background gas or plasma will appear as discrete "spicules", but do not necessarily trace out magnetic field "lines". The jets can create strong local magnetic fields due to non-overlapping electron and proton currents, consistent with the local fields observed on the solar surface and in the heliosphere. These collimated plasma jets correspond to the general concept of "flux tubes" and "flux ropes", but they have no internal "guide" field, as is commonly provided in the laboratory. As yet, no self-consistent EM description of a guide-field-free plasma jet meta-equilibrium is available.

The individual jets may interact collectively over multi-megameter distances through electric and magnetic forces, forming statistically infrequent, but stronger jets. Here, initiation and formation would be coupled to the lower boundary condition of granulations or super-granulations [48]; and to spatial sheath-charge variations; and to the upper boundary condition of return plasma flows and currents.

The accelerating jets will support the non-declining temperature of the outer Photosphere, will heat the low-density particles in the Corona, and will continue outward impeded mainly by the interaction with gas and dust, especially within about ±20° of the ecliptic plane. The solar light reflected from the energetic Jets can provide the glow of the K-Corona, thus replacing the model of a "100.eV hydrostatic gas", with the coherent energetics of a flow-through beams, reflexing beams, and infalling ecliptic gas.

Proton acceleration from the Core plasma sheath represents a major departure from "exospheric" models of the Solar Wind, where the hot Corona is both the particle source and the (thermal) energy source. That is, most modern exospheric models implicitly or explicitly assume that "The solar wind is a continuous outflow of plasma from the hot solar corona (Parker 1958)."[6] Some brief history of the two assumptions as to particle source and energetics seems warranted.

In his review chapter in *The Sun* (1953), van de Hulst [27] discussed both evidence for a 100 - 200 eV Corona; and the idea that "spicules illustrate the possibility that the chromosphere is not thermally supported." This referred to the work of Thomas [57], who modeled spicules as supersonic jets based on the hydrodynamic work of Prandtl, and mentioned (but did not analyze) "induction action arising from an electromagnetic field." Van de Hulst quantitatively discussed the proton escape velocities from Coronal radii $(1-5)R_s$, including the Pannekoek/Rosseland electric field as "classical astrophysical theory". Based on Coronal "evaporation", he estimated the total outflow of protons and electrons as $Q_{ev} \sim 10^{34.8}$ /s, somewhat lower than modern estimates of $10^{36}$ /s .

In the same volume, Kiepenheuer used sun-to-earth delay times to estimate the velocities of "two classes of corpuscular radiation" as 350 – 2000 km/s. He also develops the Biermann [7] analysis of comet tail deflections to conclude that "the sun emits a proton stream ... at all times and in all directions.", with intensity $10^{34.4}$ /s .

Parker's supersonic hydrodynamic wind [42,10] solved Eqn(4a*) with no apparent energy source except heat conduction and "constant Temperature"; and was criticized by Chamberlain [12] as mathematically invalid. However, it is still held as an exemplar of discovery.[24] Chamberlain developed a subsonic kinetic model incorporating the Pannekoek electric field, resulting in a "solar breeze" with low velocities.

The early coronal theories [3,20,56,14] determined "electron gas" density profiles from the observed light scattering, assuming $\sigma_{\gamma e} = \sigma_0$; and the required temperatures then followed as required to thermally support the plasma, informed by spectroscopic observations and thermal excitation rates. [33] Multiple models of the Solar Wind energization out of this atmosphere then followed, [17,34] leading to work on non-Maxwellian "Kappa distributions". [46,51]

More incisive analyses incorporating electric fields were developed in successive generations of "kinetic exospheric" models. [34] Second generation models with "exo base" at $r_0=6\ R_s$ produced $v_{SW} \sim 320$.km/s; incorporation of supra-thermal electrons from "kappa" distributions gave $v_{SW} \sim 450$.km/s; and more elaborate VDF analysis [32] produced velocity distribution functions with $v_{SW} \sim 600.km/s$ when the exo base was lowered to $r_0 = 1.1\ R_s$.

In contrast to exospheric models, the present electric model starts from an "electric base" in the recombination sheath at $R_s$, eliminating one entropic "atmosphere". Then, the accelerating Solar Wind *is* the K-Corona, reflecting solar light due to the same range of plasma cross-sections $\sigma_{\gamma e} >> \sigma_0$ as required to generate and maintain the photo-electric field itself. Of course, not all protons which begin acceleration below $R_s$ will reach escape velocity. The "partial launch" descriptor includes energetic protons and electrons returning sunward, adding "heat" to the coronal detritus.

A more complete electric model might also consider "runaway down electrons" [44], which are moving sunward fast enough that they are "un-correlated" with the protons; these would be "$\sigma_{\gamma e} \sim \sigma_0$ transparent" to the $\Gamma_\varepsilon$ energy "drag", but would still experience the downward electric force. This could appear as a hot "halo" distribution outside lower temperature core, similar to that observed in Helios data. [35] Further, weak anisotropic "strahl" distributions are often observed, with energies up to 600.V, sometimes with anti-sunward alignment. Recently, Scudder has developed a broad theoretical perspective on non-thermal distributions in the presence of $E_\parallel$, with comparisons of a "Steady Electron Runaway Model" to spacecraft-measured distributions. [53]

The kinetics of Solar Wind propagation continues to benefit from the excellent satellite data on particle distribution functions (PDFs) for electrons and protons, displayed in the local Wind frame $V_p$. The Helios mission down to 0.3 AU provided proton, alpha-particle, and electron PDFs, showing overall co-propagation.[35] In more detail, the protons showed "persistent skewness," with velocity deviations at the 10% $V_p$ level for perhaps 10% of the particles; and the electrons display a warm core, a somewhat hotter "halo", and a distinct "strahl" at the 1% level, in the anti-sunward direction, with energies $\sim 100.eV$. The strahl was "associated theoretically with the value of the interplanetary potential that is related to the electric field induced by gradients of the electron pressure."[35"

More modern data on the ecliptic electron PDFs has now been analyzed, from Parker Solar Probe encounters down to 0.13 AU , albeit generally from the perspective of MHD [5]. In contrast, Berčič et al. [6] analyze the observed "suprathermal electron deficit cutoff" in terms of collisionless exospheric models, to obtain the electrostatic potential $\Phi_{r\infty}$; and also analyze the observed "strahl break-point energy" in terms of a Steady Electron Runaway Model [53], to obtain a parallel electric field $E_\parallel$ in relation to the Dreicer field. This gives an electric potential $\varphi_{r\infty}(r) \sim (1.55\ kV)\ r^{-.66}$ and $E_\parallel \sim (0.84\ V/Mm)\ r^{-1.69}$ , with r scaled to $R_s$. The fractional power law signifies broadly distributed net negative charge outside the Sun, but this is not
quantified.

A separate PSP analysis by Halekas et. al [24] incorporating the same $\varphi_{r\infty}(r)$ concluded that the slow proton speed data can be explained by either exospheric models or by thermally-driven hydrodynamic models (Parker 2010) [43]; but that "neither class of model can explain the observed speed of the faster solar wind streams, which thus require additional acceleration mechanisms." [24]  Here, a model of protons accelerated from the electric-charged Sun together with resistive drag from ecliptic gas and plasma could be enlightening.

It is interesting to note that Earth magnetospheric research of the 1970's established the importance of global charge and strong potentials (~60.C, -100.kV), and strong electric fields (0.5V/m). [38,39] Models routinely invoke "double layers", [49] and definitely include the MHD-contentious "E parallel to B" even in the "conducting plasma", [19] causing strong acceleration of aurora particles.  By comparison, the electrical effects in the Sun are relatively weak, especially for such a massive structure.

The particle/fluid distinction is especially relevant for satellite measurements of the Solar Wind B-fields in the heliospheric range of 0.3 - 5. AU. [15]  There, the satellite data is broadly consistent with locally-generated magnetic fields, with pervasive random fluctuations characteristic of particles:  the magnetic energy per particle varies statistically with radius, as $B_{\mathrm{rms}}^2 / n_{\mathrm{w}} \propto \sqrt{n_{\mathrm{w}}}$ ; and "DC" values scale as the mean of random walks, so there are no persistent magnetic dipole (or monopole) fields.  The most striking collective dynamics is the occasional "dynamical arc" consistent with polarized plasma flow.  These dynamical arcs may be related to the "switchbacks" observed by PSP [5] closer to the Sun.

 Magnetic field "lines" do not glow, and a beam path does not necessarily trace a field line, especially when the beam itself creates the magnetic fields.  The conflation of energetic beam paths with magnetic field lines has contributed much to the confusion of solar models.  Indeed, the stunning Solar Orbiter movies show propagating surface flashes which bear strong resemblance to earth-surface lightning strikes in slow motion, i.e. barely slower than eV-energy electrons.  It is possible that adding electric effects to the interpretation of Solar plasmas may bring simplification and unification of the plasma dynamics and energetics.

## Acknowledgments

This work was supported by UCSD and by grants from AFOSR (FA 9550-19-1-0099 ) and DOE (DE-SC18236).  The author also acknowledges stimulating conversations with Dr. Rhon Keinigs and Dr. Jonathan Driscoll.